\documentclass[aps,pra,english,twocolumn,showpacs,preprintnumbers,amsmath,amssymb,floatfix,superscriptaddress]{revtex4-1}

\usepackage[T1]{fontenc}
\usepackage[latin9]{inputenc}
\usepackage{graphicx}
\usepackage{amssymb}
\usepackage{babel}


\usepackage{babel}
\makeatother

\begin{document}

\title{Resonance Interaction Induced by Metal Surfaces Catalyses Atom Pair  Breakage}

\author{Mathias Bostr{\"o}m}
\email{mabostr@kth.se}
\affiliation{ Department of Materials Science and Engineering, Royal Institute of Technology, SE-100 44 Stockholm, Sweden}
\affiliation{Department of Applied Mathematics, Australian National University, Canberra, Australia}

\author{Clas Persson}
\affiliation{ Department of Materials Science and Engineering, Royal Institute of Technology, SE-100 44 Stockholm, Sweden}
\affiliation{Department of Physics, University of Oslo, P. O. Box 1048 Blindern, NO-0316 Oslo, Norway}

\author{Barry W. Ninham}
\affiliation{Department of Applied Mathematics, Australian National University, Canberra, Australia}

\author{Patrick Norman}
\author{Bo E. Sernelius}
\email{bos@ifm.liu.se}
\affiliation{Division of Theory and Modeling, Department of Physics,  Chemistry and Biology, Link\"{o}ping University, SE-581 83 Link\"{o}ping, Sweden}

\begin{abstract}
We present the theory for retarded resonance interaction between two identical atoms at arbitrary positions near a metal surface. The dipole-dipole resonance interaction force that binds isotropically excited atom pairs together in free space may turn repulsive close to an ideal (totally reflecting) metal surface. On the other hand, close to an infinitely permeable surface it may turn more attractive.  We illustrate numerically how  the dipole-dipole resonance interaction between two oxygen atoms near a metal surface may provide a repulsive energy of  the same order of magnitude as the ground-state binding energy of an oxygen molecule. As a complement we also present results from density-functional theory.  
\end{abstract}


\maketitle

Two identical, isotropically excited, atoms can be bound together in free space due to attractive retarded resonance interaction. Interestingly, Jones {\it et al.} demonstrated the influence of retarded resonance interaction on the binding energy of Na$_2$ molecules\,\cite{Jones,McAlexander}. Ninham and co-workers\,\cite{Bostrom1}  demonstrated  that, due to too drastic approximations, the underlying theory of resonance interactions in free space derived from perturbative quantum electrodynamics is correct only in the non-retarded limit.
Mechanisms of breakage and formation  of molecules near surfaces are  the key problems of catalysis.
We present here a mechanism not previously noticed.  It derives from the retarded resonance interaction between two identical atoms near a surface. 
Hopmeier {\it et al.}\,\cite{Hopmeier} provided experimental evidence for enhancement of dipole-dipole interaction in  microcavities and Agarwal and
Gupta\,\cite{Agarwal} verified this theoretically. A way has recently been predicted by which very large molecules may form by resonance interaction between atoms in a narrow cavity\,\cite{BostrPRA2012}.

The aim of this brief report is to present the theory for retarded resonance interaction between two identical atoms in an excited state with the line joining the two atoms being perpendicular to the surface. A schematic illustration of the system is shown in Fig.\,\ref{figu1}.  This choice of configuration was used because it enables us to derive analytical results for a  configuration that has not been previously studied. Previous work considered only the case with two atoms at the same distance from the surfaces. This is now extended to enable consideration of two atoms at arbitrary positions near a metal surface.  We find that attractive resonance interaction between isotropically excited atom pairs in free space may turn repulsive close to an ideal metal surface and more attractive close to an infinitely permeable surface. Close to an ideal metal there is a repulsive contribution which decreases as $1/z_+^3$ (where $z_+$ is the distance between one atom and the surface image of the second atom).  The calculations of resonance interaction between excited-state atom pairs and Casimir-Polder interaction between ground-state atom pairs require the proper  Green's functions. The non-zero matrix elements of the field susceptibility matrix ($T_{ii}$) needed for our calculations were adapted from the Green's functions given in the literature. 
Buhmann and co-workers\,\cite{Buhmann1,Buhmann2} studied Casimir-Polder interaction between two ground-state atoms placed with different orientations near a surface.  As an illustrative example we present numerical calculations on the resonance interaction between two oxygen atoms near a metal surface.  The bond enthalpy for an O$_2$ molecule  is around 498 kJ mol$^{-1}=5.2$ eV. The sun-light spectrum has an energy range of 1--3 eV, so apart from solar energy one needs an  extra 3 eV to split an oxygen molecule. A contribution to such bond splitting may occur due to excitation-induced resonance interaction between oxygen atoms near a metal surface. We also present results for two Zn atoms where the binding energy is much smaller, $ \sim 0.4$ eV, and the resonance interaction effects are much stronger.

\begin{figure}
\includegraphics[width=8cm]{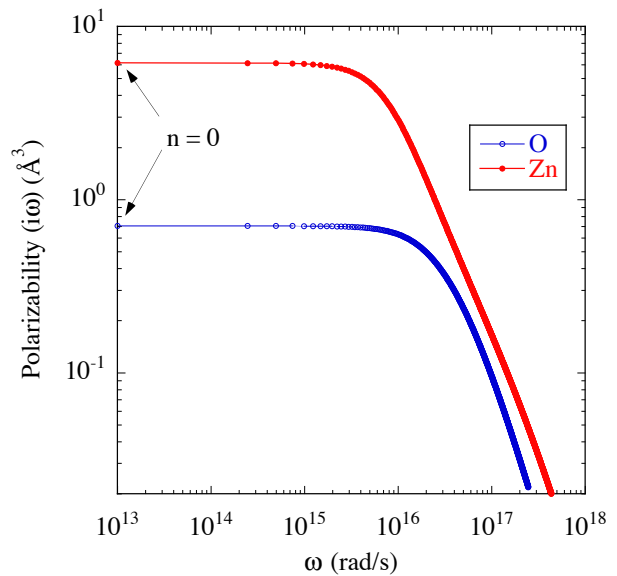}
\caption{(Color online) The polarizabilities of oxygen and zinc atoms. The discrete Matsubara frequencies are indicated by the circles. Note that we have placed the $n=0$ value at the left vertical axes. }
\label{figu2}
\end{figure}

We first recall the standard argument:  Consider two identical atoms where one initially is in its ground state and the other is in an excited state.  This state can also be represented by a superposition of states, one symmetric and one antisymmetric with respect to interchange of the atoms. While the symmetric state is likely to decay into two ground-state atoms, the antisymmetric state can be quite long-lived. The system can thus be trapped in the antisymmetric state\,\cite{Bostrom1,Stephen}. The energy migrates back and forth between the two atoms until either the two atoms move apart or a photon is emitted away from the system. First order dispersion interactions are caused by this coupling of the system, i.e. the energy difference between the two states is separation ($\rho$) dependent. After writing down the equations of motion for the excited system it is straightforward to derive the zero temperature Green's function for two identical and isotropic atoms\,\cite{McLachlan,Bostrom1}. The  resonance frequencies of the system are given by the following equation\,\cite{Bostrom1}:

\begin{figure}
\includegraphics[width=6cm]{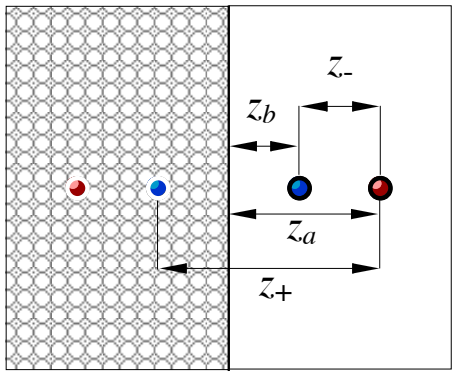}
\caption{(Color online) Two atoms in vacuum near a metal surface. Also shown are their mirror images}
\label{figu1}
\end{figure}

\begin{equation}
1-\alpha(1|\omega) \alpha(2|\omega) T(\rho|\omega)^2=0,
\label{Eq1}
\end{equation} 
where $\alpha(j|\omega)$ is the polarizability of atom $j$\,\cite{London}. We present in  Fig.\,\ref{figu2} the numerically evaluated polarizability of oxygen and zinc atoms used in this study.  The atomic polarizability data were obtained from quantum chemical response theory calculations based on the complex polarization propagator (CPP) approach\,\cite{Nor} with implementations made in the Dalton program\,\cite{Dalt} as described in Refs. \cite{Nor2,Nor3,Kau}. In the CPP framework, damping terms are introduced in the response functions so as to describe relaxation in the system. These damping terms may be viewed as causing the frequency argument of the linear response function to become complex, and in the present work we employ this feature to determine the polarizability for the special case of imaginary frequencies. We have earlier described and applied this technique in several papers, see, e.g., Ref.\,\cite{Nor4}.

The electronic configuration of the Zn atom is [Ar]$3d^{10}4s^{2}$ and we chose to describe the electronic structure of this singlet ground state at the Kohn-Sham density functional theory (DFT) level with use of the Coulomb attenuated hybrid exchange-correlation functional CAM-B3LYP\,\cite{Becke,Yanai} in conjunction with Dunning's triple-augmented triple-$\zeta$ basis set (t-aug-cc-pVTZ) \cite{Bala}.  For the O atom we consider the reference state of triplet spin symmetry and we describe the wave function by means of the multiconfigurational self- consistent field (MCSCF) method. We apply a valence active space, which means there is only one triplet spin-adapted configuration with $M_S=0$. Also in this case, we adopt the t-aug-cc-pVTZ basis set\,\cite{Ken}.

In the case of two identical atoms the above resonance condition can be separated in one antisymmetric and one symmetric part. Since the excited symmetric state has a much shorter life time than the antisymmetric state, the system can be trapped in an excited antisymmetric state\,\cite{Bostrom1}.  The resonance interaction energy of this antisymmetric state is,
\begin{equation}
\label{Eq2}
U(\rho)= \hbar [\omega_{r} (\rho)-\omega_{r} (\infty)].
\end{equation}
Since the relevant solution of  Eq.\,(\ref{Eq1}) really is the pole of the antisymmetric part of the underlying Green's function we can in a standard way\,\cite{Sernelius} deform a contour of integration around this pole to obtain a both simple and exact expression for the resonance interaction energy,
\begin{equation}
\begin{array}{l}
U(\rho ) = \sum\limits_{i = x,y,z} {\frac{\hbar }{\pi }\int_0^\infty  d \omega ln\left[ {1 + \alpha (1|i\omega ){T_{ii}}(\rho |i\omega )} \right]} \\
\quad \quad \quad  \approx \sum\limits_{i = x,y,z} {\frac{\hbar }{\pi }\int_0^\infty  d \omega \alpha (1|i\omega ){T_{ii}}(\rho |i\omega )} .
\end{array}
\label{Eq3}
\end{equation}

In order to compare with previous calculations on dipole-dipole resonance interaction we use the approximate linearized expression. In this way we ignore corrections due to an attractive van der Waals (Casimir-Polder) term. We also ignore finite size effects which will change the interaction for separations smaller than two atomic radii. 

To account for the temperature ($T$) dependence we simply replace the integration over imaginary frequencies by a summation over discrete Matsubara frequencies\,\cite{Sernelius},
\begin{equation}
\frac{\hbar }{{2\pi }}\int_0^\infty  d \omega  \to {k_B}T\sum\limits_{n = 0}^\infty  {\rm{'}} ,\quad {\omega _n} = 2\pi {k_B}Tn/\hbar ,
\label{Eq4}
\end{equation}
where $k_B$ is the Boltzmann constant and the prime indicates that the $n=0$ term should be divided by 2. 

\begin{figure}
\includegraphics[width=8cm]{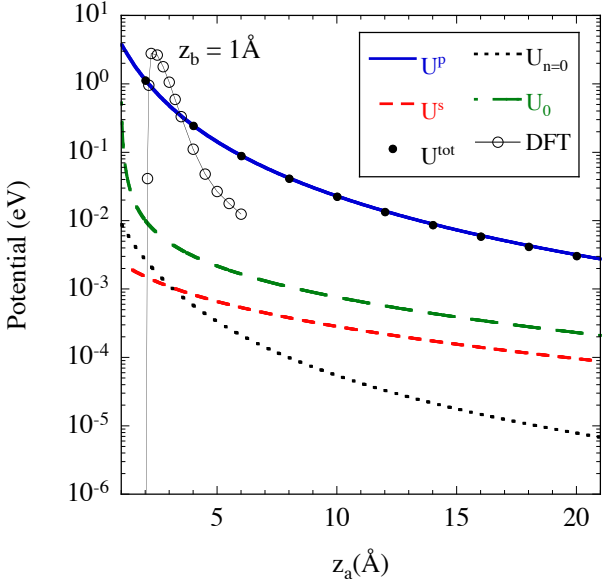}
\caption{(Color online) The resonance interaction energy between two oxygen atoms situated one outside the other near an ideal metal surface. The closest atom is at 1 {\AA} from the surface. For comparison we have added a curve (DFT-curve) showing the potential for two oxygen atoms at an Au (111) surface as obtained from a DFT calculation including a van der Waals functional. Se the text for details}
\label{figu3}
\end{figure}

\begin{figure}
\includegraphics[width=8cm]{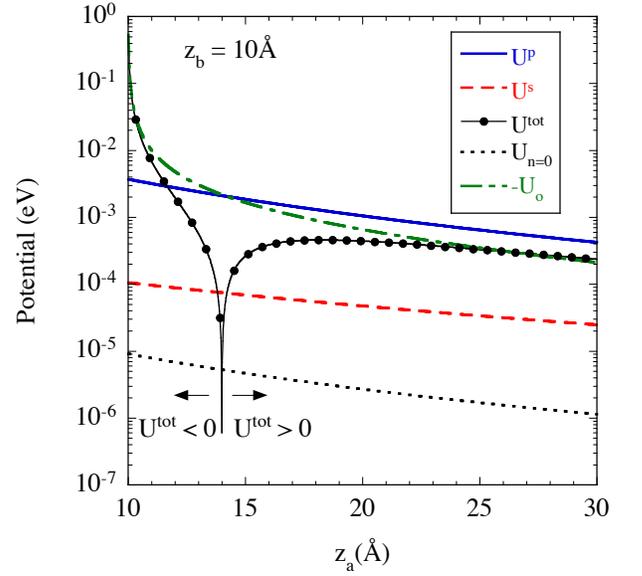}
\caption{(Color online) The resonance interaction energy between two oxygen atoms situated one outside the other near an ideal metal surface. The closest atom is 10 {\AA} from the surface.}
\label{figu4}
\end{figure}

We consider the case when the distance, $x$, between the two atoms in the plane of the surface is zero and the distances from atoms $a$ and $b$ to the surface is $z_a$ and $z_b$, respectively. In other words the atoms are along a line perpendicular to the surface; atom $a$ is furthest away from the surface (see Fig.\,\ref{figu1}).
The field susceptibility is a sum of three terms. Apart from the free space field susceptibility  $ T_{ii}^0$, there are $p$ and $s$  field susceptibility corrections due to the presence of a surface,  $T_{ii}^p$ and $T_{ii}^s$, respectively. In free space the field susceptibility matrix ${\bf  T}(\rho|i \omega)$ has the following non-zero matrix elements:

\begin{equation}
\begin{array}{l}
T_{xx}^0(i{\omega _n}) = T_{yy}^0(i{\omega _n}) =- (\frac{{\omega _n^2}}{{{c^2}}} + \frac{{{\omega _n}}}{{c{z_ - }}} + \frac{1}{{z_ - ^2}})\frac{{{e^{ - \omega {z_ - }/c}}}}{{{z_ - }}},\\
T_{zz}^0(i{\omega _n}) =  2(\frac{1}{{z_ - ^2}} + \frac{{{\omega _n}}}{{c{z_ - }}})\frac{{{e^{ - {\omega _n}{z_ - }/c}}}}{{{z_ - }}}.
\end{array}
\label{Eq5}
\end{equation}
Here we define $z_+=z_a+z_b$ and $z_-=z_a-z_b$, with the first being the distance between atom $a$ and the image of atom $b$ inside the surface and the second being the distance between the two atoms (see Fig.\,\ref{figu1}). The corresponding surface-induced corrections to the field susceptibility matrix \,\cite{Buhmann1,Buhmann2} have contributions from $p$ and $s$ polarizations (with $\eta=\omega_n z_+/c$),

\begin{equation}
\begin{array}{l}
T_{xx}^p = T_{yy}^p =  {r_p}[1 + \eta  + 0.5{\eta ^2}]{e^{ - \eta }}/z_ + ^3,\\
T_{zz}^p =   2{r_p}(1 + \eta ){e^{ - \eta }}/z_ + ^3,
\end{array}
\label{Eq6}
\end{equation}
and
\begin{equation}
\begin{array}{l}
T_{xx}^s = T_{yy}^s = {-r_s}\omega _n^2{e^{ - \eta }}/(2{c^2}{z_ + }),\\
T_{zz}^s = 0,
\end{array}
\label{Eq7}
\end{equation}
respectively.
Here $r_s$ and $r_p$ are the reflection coefficients for $s$- and $p$-polarized waves, respectively, impinging on the surface. For an ideal metal $r_s=-1$ and $r_p=1$. For an infinitely permeable surface one gets a sign reversal of both terms. Therefore the corrections to the resonance interaction due to a surface have opposite sign at an ideal metal surface as compared to an infinitely permeable surface. The result in free space in the retarded limit  is 

 \begin{equation}
U(z_-) \approx-2 \hbar c \alpha (0)/(\pi z_-^4).
\label{Eq8}
\end{equation}

Near a surface we find a surface-induced  nonretarded resonance interaction between the two atoms,

\begin{figure}
\includegraphics[width=8cm]{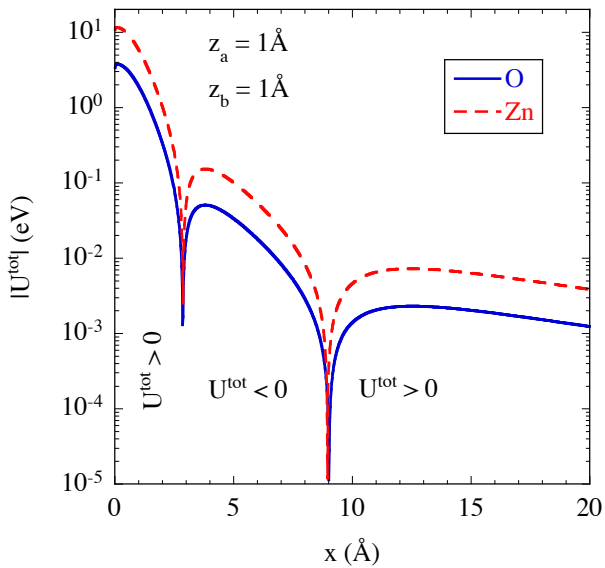}
\caption{(Color online) The resonance energy between two oxygen (zinc) atoms both adsorbed on an ideal metal surface ($z_a=z_b=$1 \AA), solid (dashed) curve. The field susceptibility for this case is not described in the text.  The different contributions can be obtained using the Green's functions given in the literature\,\cite{Buhmann1,Buhmann2}.}
\label{figu5}
\end{figure}

 \begin{equation}
U(z_+) \approx4 \hbar r_p/(\pi z_+^3) \int_0^\infty d \omega \alpha (i \omega) \approx 2  r_p \hbar \omega_j \alpha (0)/(z_+^3),
\label{Eq9}
\end{equation}
where the London approximation, $\alpha (j|i\omega ) \approx \alpha (j|0)/\left( {1 + {\omega ^2}/\omega _j^2} \right)$, has been used to find the final analytical expression. It is also possible to derive analytical results for the surface corrections to the resonance interaction at finite temperature in the retarded limit when $\alpha (i \omega_n) \approx \alpha (0)$. Defining $\omega_n=n \omega_1$,  $\eta_1=\omega_1 z_+/c$, and $\gamma=(-1+e^{\eta_1})^{-1}$,  we find for the $p$- and $s$-polarized contributions,

 \begin{equation}
\begin{array}{l}
{U^p}({z_ + })  \approx \frac{{2{r_p}{k_B}T\alpha (0)}}{{z_ + ^3}}\left\{ {4 + 2\eta _1^2{\gamma ^3}} \right.\\
\quad \quad \quad \quad  + \left. {(4 + 4{\eta _1} + \eta _1^2)\gamma  + (4{\eta _1} + 3\eta _1^2){\gamma ^2}} \right\},
\end{array}
\label{Eq10}
\end{equation}
and
 \begin{equation}
{U^s}({z_ + }) \approx \frac{{ - 2{r_s}{k_B}T\alpha (0)\eta _1^2}}{{{z_ + }}}\{ 2{\gamma ^3} + 3{\gamma ^2} + \gamma \},
\label{Eq11}
\end{equation}
respectively. The retarded asymptote at zero temperature then is
 \begin{equation}
U^s (z_+)+U^p (z_+) \approx\hbar c \alpha (0)   (10 r_p-2 r_s)  /(\pi z_+^4).
\label{Eq12}
\end{equation}
It is interesting to see that this asymptote has the same power law as the Casimir-Polder energy between a ground-state atom and an ideal metal surface\,\cite{Casimir}, which goes as $U^{CP} (z) \approx-3 \hbar c \alpha (0)   /(8 \pi z^4)$.

At finite temperature the long range resonance interaction driven by entropy is given by

 \begin{equation}
U^s (z_+)+U^p (z_+) \approx 8 r_p k_B T \alpha (0)/z_+^3.
\label{Eq13}
\end{equation}

Now let us consider the total resonance interaction in the retarded limit but at sufficiently small separations for the zero-temperature results to apply. It is given by

 \begin{equation}
U \approx [\hbar c \alpha (0)/\pi] \{ [ (10 r_p-2 r_s)  /(z_+^4)]-2/z_-^4\}.
\label{Eq14}
\end{equation}

It is then clear that in the case of two atoms near an ideal metal surface the total resonance interaction can be either attractive or repulsive depending on $z_+$ and $z_-$. 

We will now proceed to calculate numerically the retarded resonance interaction at 300 K between two oxygen atoms near an ideal metal surface. We present in Figs.\,\ref{figu3} and\,\ref{figu4}  the retarded resonance interaction between two atoms near an ideal metal surface.  The total potential is the sum of the contributions from free space $U_0$, the correction due to the surface from $s$ waves $U^s$, and the correction due to the surface from $p$ waves  $U^p$. In the first case, when one atom is adsorbed at the surface,  the interaction is dominated by $p$-polarized surface corrections. When the atom closest to the surface is some distance away from the surface the interaction for small atom-atom separations is dominated by the free space contribution. However, as the distance between the atoms increases repulsive surface corrections become increasingly important, as can be seen in Fig.\,\ref{figu4}. In Fig.\,\ref{figu3} we also compare the total potential with the potential energy for two oxygen atoms on a gold surface calculated within the Kohn-Sham DFT model with a nonlocal van der Waals correlation functional (rPW86/vdw-DF2)\,\cite{Kre,Kli}. In the atomistic calculation the Au (111) surface is modeled by a 6x6x6
supercell slab structure with 44 {\AA} vacuum layer. The fixed oxygen atom is relaxed at the fcc hollow surface site, and the position of the second atom is varied in the $z$ direction. We find that the non-resonant contribution to the interaction affects the result for very small separations of the order of 3 {\AA} or less. We have in recent work found that this is also the  limit where finite-size corrections influence the interaction. Our results are therefore only meaningful  when the two atoms are at separations beyond close contact.

Extending the theory presented above, we furthermore calculated the retarded resonance interaction between two atoms both adsorbed on an ideal metal surface. We show the result in Fig.\,\ref{figu5} as a function of the lateral distance $x$ between the two atoms.  Surface corrections to the resonance interaction between atoms near an ideal metal surface may result in a significant repulsion. The DFT calculations of zinc and oxygen on the gold surface reveal that these atom pairs have a very weak chemical bond when both atoms are adsorbed on the true gold surface.

We have discussed how the presence of a surface influences  the retarded dipole-dipole resonance interaction between two identical atoms in an isotropically excited state. For very small atom-atom separations, close to direct contact, there will be important finite-size corrections and influence from non-resonant contributions  such as non-retarded van der Waals forces. Beyond these limiting separations  the excitation-induced interaction between two atoms near an ideal metal can produce a repulsive energy of the same order of magnitude as the bonding energy of diatomic molecules. Catalytic effects due to the presence of a surface may induce large enough energies to break up or form molecules.  The  ${\rm{O}_2}$ reduction reaction is one of the most studied, not only because it is interesting scientifically but also important for practical purposes. Such catalytic effects occur, for example, at the cathode electrode of fuel cells, where Pt, Au, and related alloys have been used
or suggested as the catalyst. Yotsuhashi {\it et al.}\,\cite{ Yotsuhashi} studied dissociative adsorption of ${\rm{O}_2}$ molecules on Pt and Au surfaces. They found that the chemical reactivity is substantially larger on Pt as compared to on Au surfaces.  Another interesting surface is Cu(110) where the energy gain by dissociation of an ${\rm{O}_2}$ molecule on the surface is approximately 1.3 eV per molecule\,\cite{Liem}. There are several diatomic molecules and metallic surfaces
that will show varying strengths of the chemical interactions.
For instance, we observe that the O$_2$ molecule has
much larger bond enthalpy  in vacuum ($\sim$5 eV) compared to the
bond enthalpy of the Zn$_2$ molecule ($\sim$0.3 eV).
Also, we regard Au to be somewhat more comparable to
an ideal metal than for example, Pt, and we therefore
complement the calculation of resonance interactions with
atomistic modeling of O$_2$ on Au(111) surfaces.

%
M.B. and C.P. acknowledge support from VR (Contract No. C0485101) and STEM (Contract No. 34138-1).
B.E.S. also acknowledges financial support from VR (Contract No. 70529001). 

\end{document}